\documentclass[%
 reprint,
superscriptaddress,
 amsmath,amssymb,
 aps,
prl,
]{revtex4-2}

\usepackage{graphicx}
\usepackage{dcolumn}
\usepackage{bm}

\usepackage{upgreek}

\usepackage{xcolor}

\begin{document}



\title{Chiral Phonons with Giant Magnetic Moments in 
a Topological Crystalline Insulator}

\author{Felix G.\ G.\ Hernandez}
\thanks{These authors contributed equally to this work}
\affiliation{Instituto de F\'{i}sica, Universidade de S\~{a}o Paulo, S\~{a}o Paulo, SP 05508-090, Brazil}
\email{felixggh@if.usp.br}

\author{Andrey Baydin}
\thanks{These authors contributed equally to this work}
\affiliation{Department of Electrical and Computer Engineering, Rice University, Houston, Texas 77005, USA}%
\affiliation{Smalley-Curl Institute, Rice University, Houston, Texas, 77005, USA}

\author{Swati Chaudhary}
\affiliation{Department of Physics, The University of Texas at Austin, Austin, Texas 78712, USA}
\affiliation{Department of Physics, Northeastern University, Boston, Massachusetts 02115, USA}
\affiliation{Department of Physics, Massachusetts Institute of Technology, Cambridge, Massachusetts 02139, USA}

\author{Fuyang Tay}
\affiliation{Department of Electrical and Computer Engineering, Rice University, Houston, Texas 77005, USA}
\affiliation{Applied Physics Graduate Program, Smalley-Curl Institute, Rice University, Houston, Texas, 77005, USA}

\author{Ikufumi~Katayama}
\affiliation{Department of Physics, Graduate School of Engineering Science, Yokohama National University, Yokohama 240-8501, Japan}

\author{Jun Takeda}
\affiliation{Department of Physics, Graduate School of Engineering Science, Yokohama National University, Yokohama 240-8501, Japan}

\author{Hiroyuki Nojiri}
\affiliation{Institute for Materials Research, Tohoku University, Sendai 980-8577, Japan}

\author{Anderson K.\ Okazaki}
\affiliation{Instituto Nacional de Pesquisas Espaciais, S\~{a}o Jos\'{e} dos Campos, SP 12201-970, Brazil}

\author{Paulo H.\ O.\ Rappl}
\affiliation{Instituto Nacional de Pesquisas Espaciais, S\~{a}o Jos\'{e} dos Campos, SP 12201-970, Brazil}

\author{Eduardo~Abramof}
\affiliation{Instituto Nacional de Pesquisas Espaciais, S\~{a}o Jos\'{e} dos Campos, SP 12201-970, Brazil}

\author{Martin Rodriguez-Vega}
\affiliation{Department of Physics, The University of Texas at Austin, Austin, Texas 78712, USA}
\affiliation{Department of Physics, Northeastern University, Boston, Massachusetts 02115, USA}

\author{Gregory A.\ Fiete}
\affiliation{Department of Physics, Northeastern University, Boston, Massachusetts 02115, USA}
\affiliation{Department of Physics, Massachusetts Institute of Technology, Cambridge, Massachusetts 02139, USA}

\author{Junichiro Kono}
\affiliation{Department of Electrical and Computer Engineering, Rice University, Houston, Texas 77005, USA}%
\affiliation{Smalley-Curl Institute, Rice University, Houston, Texas, 77005, USA}
\affiliation{Department of Physics and Astronomy, Rice University, Houston, Texas 77005, USA}%
\affiliation{Department of Material Science and NanoEngineering, Rice University, Houston, Texas 77005, USA}

\date{\today}

\begin{abstract}
We have studied the magnetic response of transverse optical phonons in Pb$_{1-x}$Sn$_{x}$Te films. 
Polarization-dependent terahertz magnetospectroscopy measurements revealed Zeeman splittings and diamagnetic shifts, demonstrating that these phonon modes become chiral in magnetic fields.  
Films in the topological crystalline insulator phase ($x > 0.32$) exhibited magnetic moment values that are larger than those for topologically trivial films ($x < 0.32$) by two orders of magnitude. Furthermore, the sign of the effective $g$-factor was opposite in the two phases, a signature of the topological transition within our model.  These results strongly indicate the existence of interplay between the magnetic properties of chiral phonons and the topology of electronic band structure.
\end{abstract}


\maketitle

The symmetries of crystals determine many properties of their phonons, such as their selection rules and degeneracies~\cite{RevModPhys.40.1}. Among the possible symmetries, mirror symmetries play a special role: when they are broken, the lattice ions can display circular motion with finite angular momentum. These modes are called chiral phonons~\cite{PhysRevLett.115.115502,doi:10.1126/science.aar2711,PhysRevResearch.4.013129, PhysRevB.105.184412,PhysRevB.106.144302}. In magnetic fields, chiral phonons preferably absorb polarized light of a given handedness, resulting in magnetic circular dichroism (MCD)~\cite{ANASTASSAKIS1971563,ANASTASSAKIS19721091,PhysRevLett.128.075901}. Furthermore, chiral phonons carry a finite magnetic moment that induces a phonon Zeeman effect, which has been observed in the narrow-gap semiconductor PbTe~\cite{PhysRevLett.128.075901}, 
the non-axial CeF$_3$~\cite{Schaack_1976}, and the Dirac semimetal Cd$_3$As$_2$~\cite{ChengetAl2020NL}, with magnetic moment values ranging from hundredths to several Bohr magnetons.

 
There are a few mechanisms that can lead to a finite phonon magnetic moment. Recent theoretical reports~\cite{JuraschekEtAl2017PRM,JuraschekSpaldin2019PRM} have proposed that a phonon magnetic moment can arise from the time-dependent electric polarization induced by a laser-driven infrared active phonon. However, the predicted magnitude of the phonon magnetic moment, which depends on the phonon effective charge and the ion masses, is relatively small compared to recent experimental observations~\cite{PhysRevLett.128.075901,ChengetAl2020NL}. Larger values compatible with experiments can be obtained when electronic contributions are considered. In this regard, Ref.~\cite{ren2021phonon} proposed a mechanism where the circular motion of a chiral phonon induces an electronic orbital response that contributes to the phonon magnetic moment. Further, Ref.~\cite{PhysRevResearch.4.L012004} suggested that the chiral phonon can induce inertial effects on the electrons, which lead to an effective spin-chiral phonon coupling. Therefore, electronic contributions to the phonon magnetic moment open the possibility for the interplay of chiral phonons and electronic topology~\cite{ren2021phonon}. However, no experimental evidence has been reported. 




\begin{figure*}[ht!]
 \centering
 \includegraphics[width=1\textwidth]{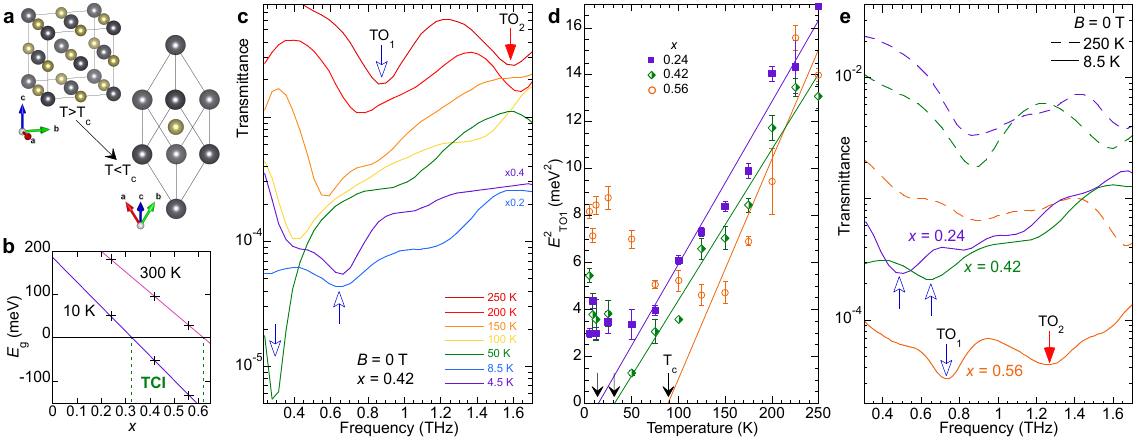}
 \caption{(a)~Cubic and distorted unit cell structure of Pb$_{1-x}$Sn$_{x}$Te, where gray spheres represent Pb/Sn atoms and yellow spheres represent Te atoms. (b)~Diagram of the trivial-to-TCI transition in the 10--300\,K range, adapted from Ref.~\cite{PhysRevLett.16.1193}. The crosses mark our samples. (c)~Transmittance spectra for the $x=0.42$ sample 
 at various temperatures. (d)~Temperature dependence of the energy squared of the TO$_1$ phonon mode for all samples. The solid lines are Curie-Weiss fits. (e)~Transmittance spectra 
 above and below $T_\text{c}$ for all samples. $B=0$\,T.
 }
 \label{fig:1}
\end{figure*}


In this Letter, we report results of terahertz time-domain spectroscopy (THz-TDS) experiments on the pseudobinary alloy Pb$_{1-x}$Sn$_{x}$Te. This material system is known to be a topological crystalline insulator (TCI) for $x>0.32$~\cite{Hsieh_2012,Tanaka2012} and has exhibited chiral phonons in strong magnetic fields for $x=0$~\cite{PhysRevLett.128.075901}. We studied one sample in the trivial phase ($x=0.24$) and two samples in the topological phase ($x= 0.42$ and $0.56$). For each sample, we observed two anharmonicity-split transverse optical (TO) phonon modes and characterized their magnetic properties at temperatures low-enough to place the samples in their ferroelectric phases~\cite{PhysRevLett.16.1193,KAWAMURA1975341}. Both the trivial and topological samples exhibited chiral phonons. However, across the topological transition, the phonons switched chirality, and the phonon magnetic moment increased by two orders of magnitude. We supplemented our experimental observations with a theoretical model for the phonon magnetic moment arising from the electronic orbital response~\cite{ren2021phonon}, which captures the chirality switch across the topological transition. Thus, our results indicate a connection between electronic topology and phonon chirality.

The single-crystal Pb$_{1-x}$Sn$_x$Te films ($\sim$1.5~$\upmu$m thick) were epitaxially grown on (111) BaF$_2$ substrates. This alloy system is known to exhibit $x$-dependent ferroelectric and topological phase transitions~\cite{PhysRevLett.16.1193,KAWAMURA1975341}. The existence of the ferroelectric phase is the result of a rhombohedral distortion of the rock-salt structure, which breaks inversion symmetry and leads to a ferroelectric transition at a critical temperature ($T_\text{c}$)~\cite{PhysRevLett.37.772,Burkhard:77,SUGAI1977127,PhysRevLett.112.175501,NatComm7.12291,Ribeiro2018,PhysRevResearch.2.012048,PhysRevB.102.024112,PhysRevB.102.115204,PhysRevB.102.155132}; see Fig.~\ref{fig:1}(a). Additionally, with increasing $x$, the electronic band gap ($E_\text{g}$) decreases until it closes at a critical value ($x_\text{c}$) and the electronic bands invert around the L-points~\cite{PhysRevLett.16.1193,LENT1986491}, entering the TCI state~\cite{PhysRevB.90.235114,li2016,Varjas2020,Okazaki2018,ChengetAl2019PRL,liu2013,safaei2013,Assaf2016,Xing2008,Xing2008,Xu2012}. The critical concentration $x_\text{c}$ is temperature-dependent~\cite{PhysRevLett.16.1193}; see Fig.~\ref{fig:1}(b).

We have performed standard THz-TDS measurements in transmission geometry in high magnetic fields~\cite{Baydinetal2021FO,SM}. 
First, to identify the spectral features corresponding to TO phonons, we report measurements at various temperatures at zero magnetic field. Transmittance spectra for the $x=0.42$ sample are shown in Fig.~\ref{fig:1}(c). The minima observed at $\sim$0.9\,THz (TO$_1$) and $\sim$1.6\,THz (TO$_2$) at high temperatures are consistent with two anharmonicity-split optical phonon modes observed in neutron scattering measurements on PbTe~\cite{DelaireetAl2011NM} and similar to features observed in Pb$_{1-x}$Sn$_x$Se~\cite{doi:10.1021/acsphotonics.1c01717}. The large widths of the phonon resonances are related to alloy-inherent disorder. The phonon modes red-shift, or soften, continuously with decreasing temperature until the temperature reaches $T_\text{c}$. Below $T_\text{c}$, the modes blue-shift (or harden) with decreasing temperature but less steep than expected by the Landau theory of ferroelectrics~\cite{PhysRevB.95.144101,Kittel}, as also reported in materials with degenerate soft phonons~\cite{RevModPhys.46.83}.
Similar $\Gamma$-point TO phonon behavior has been associated with a ferroelectric transition of the displacive nature for Pb$_{1-x}$Sn$_{x}$Te~\cite{npj2022}, and SnTe~\cite{KOBAYASHI1975875,PhysRevB.95.144101}. 


To characterize the ferroelectric phase in our samples, we fit the transmittance spectra modelling TO phonons as Lorentzians~\cite{SM}. 
Increasing from 15\,K up to 89\,K, $T_\text{c}$ was obtained from Curie-Weiss fits in the phonon softening region of Fig.~\ref{fig:1}(d).
We note that the TO phonons in the $x=0.24$ and $0.42$ samples soften to a lower energy than in the $0.56$ sample. This behavior can be related to the interplay between $T_\text{c}$ and the density of free carriers~\cite{PhysRevLett.17.753}, which increase with $x$ because the native Sn vacancies make Pb$_{1-x}$Sn$_{x}$Te epilayers more $p$-type~\cite{doi:10.1063/1.366051,SM}. 

The phonon frequency hardening in Fig.~\ref{fig:1}(d) shows that all the studied films were in the ferroelectric phase below 15\,K.
Transmittance spectra for the samples 
are shown in Fig.~\ref{fig:1}(e) above and below $T_\text{c}$. 
In the last case, the phonon angular momentum, $J^{\mathrm{ph}}=\sum_{\alpha} m_{\alpha} \boldsymbol{u}_{\alpha} \times \partial_t \boldsymbol{u}_{\alpha}$, can be nonzero for such crystal without inversion symmetry~\cite{PhysRevLett.115.115502}. Here, the index $\alpha$ runs over all the atoms in the crystal, $\boldsymbol{u}_{\alpha}$ is the phonon displacement vector, and $m_{\alpha}$ is the phonon mass. This low temperature condition is maintained in the following discussion allowing to host chiral phonons in all the samples and to access the TCI phase above $x_\text{c} \sim 0.32$. 

\begin{figure}[h]
 \centering
 \includegraphics[width=1.0\columnwidth]{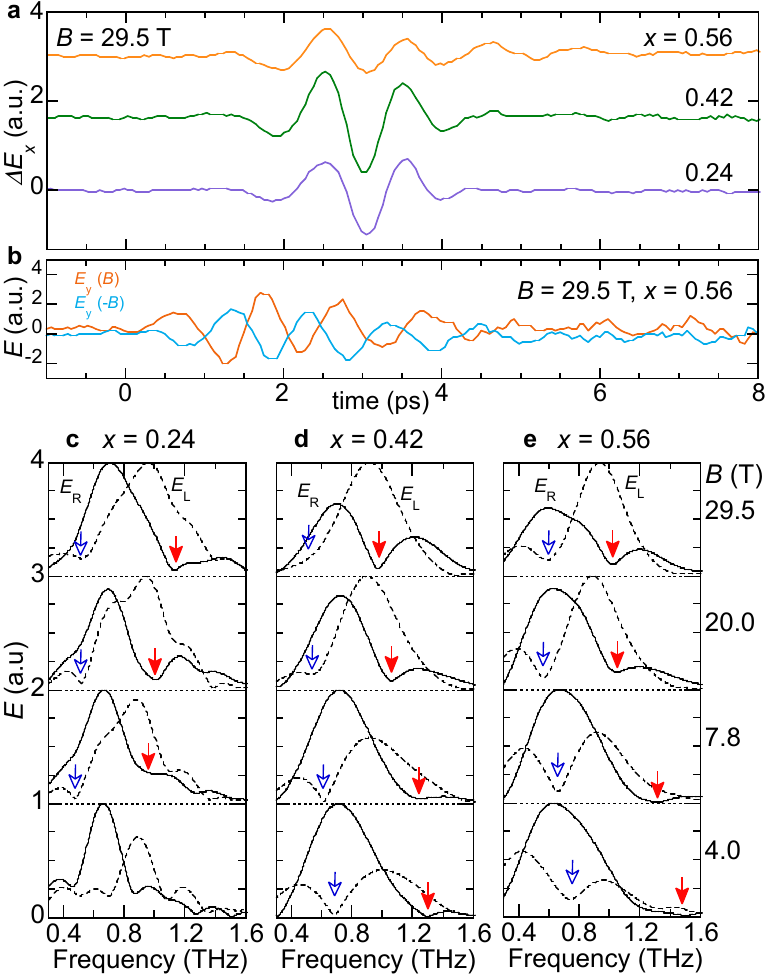}
 \caption{
 (a)~Magnetic-field-induced change in $E_x$ at 29.5\,T for all samples. (b)~Magnetic-field-induced $E_y$ component for $x=0.56$ at $\pm$29.5\,T. (c)-(e)~Right (solid line) and left-handed (dashed line) transmitted electric fields in the frequency domain for several $B$. The data was normalized. The full and empty arrows indicate the position of relevant transmission minima in $E_\text{R,L}$ as a guide to the eye for the circular dichroism. Scans in (a) and (c)-(e) are vertically offset. $T=12$\,K. }
 \label{fig:2}
\end{figure}

Next, to investigate the magnetic properties of these TO phonons across the topological transition, we applied magnetic field pulses up to 30\,T synchronized with THz radiation~\cite{NoeetAl2016OE,Baydinetal2021FO,SM}. While the incident polarization was set in the x direction, we measured both linear components of the transmitted THz electric fields with a detection polarizer oriented along the $x$ ($E_x$) or $y$ directions ($E_y$). The magnetic-field-induced changes at 29.5\,T are displayed in Figs.~\ref{fig:2}(a) and \ref{fig:2}(b). The change in the $E_x$ component ($\Delta E_x$) can be clearly seen to contain several oscillations in time, with the oscillation amplitude slightly reduced for the $x=0.56$ sample. 
The magnetic-field-induced $y$-component, $E_y$, is robust and has a few picosecond oscillations that can be reversed by changing the polarity of the magnetic field ($B$).  An $E_y$ component with amplitude comparable to $E_x$ is a clear indication of the existence of phonon chirality, for which a circular right(R)/left(L) basis, $E_\text{R,L}=(E_x\pm iE_y)/\sqrt{2}$, is more adequate for analysis. 

Figures~\ref{fig:2}(c)-\ref{fig:2}(e) show $E_\text{R,L}$ in the frequency domain for all samples for different $B$. The empty (blue) and solid (red) arrows indicate the positions of the transmission minima for the opposite senses of circular polarization. A high degree of MCD is obtained when a transmission minimum occurs for one polarization while the opposite one displays a maximum. The plots show significant differences in $E_\text{R,L}$, depending on whether the sample is a trivial insulator or a TCI. Before the topological transition, at $x=0.24$, the data has several wiggles at low magnetic fields whose origin cannot be identified. Magnetic fields larger than 7.8\,T are required to observe a deep minimum in $E_\text{R,L}$. On the other hand, on the TCI side ($x=0.42$ and $0.56$) $E_\text{R,L}$ has well-defined deep minima clearly visible for low fields. This implies that a larger degree of MCD can be obtained in the TCI phase, reaching nearly 100\% at a much lower $B$ than that needed for the trivial insulator. Note that the MCD value we recently reported for TO phonons in PbTe ($x=0$) at 9\,T was 30\%~\cite{PhysRevLett.128.075901}.

\begin{figure*}[ht!]
 \centering
 \includegraphics[width=1.0\textwidth]{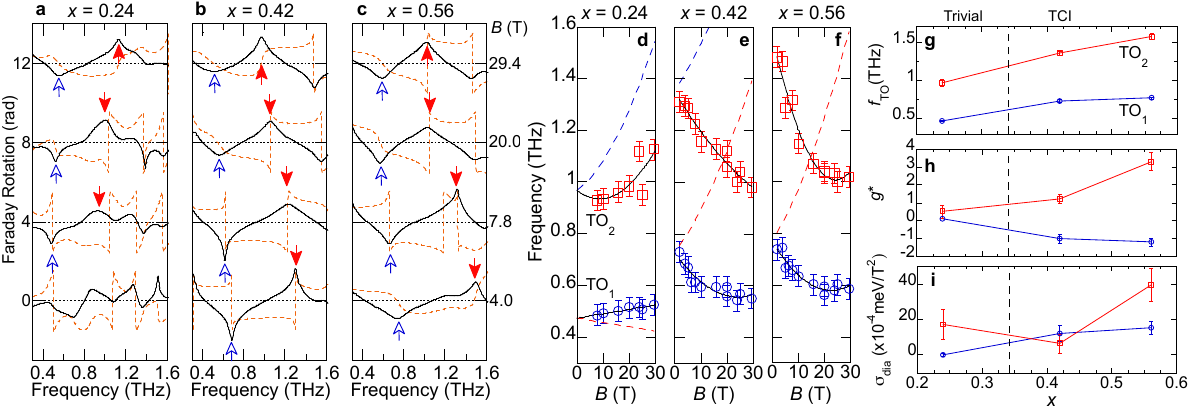}
 \caption{(a)-(c)~$\Theta$ (dashed lines) and $\eta$ (solid lines) as a function of frequency for several $B$. The arrows indicate the frequencies shown in Fig.~\ref{fig:2}. (d)-(f)~Frequency of the ellipticity peaks as a function of $B$ with fitting model (solid lines). The dashed lines estimate the position of phonon branches with complementary chirality. The obtained parameters are plotted in (g-i). Scans shown in (a)-(c) are vertically offset for clarity. $T=12$\,K.
 }
 \label{fig:3}
\end{figure*}


The ferroelectric transition could be responsible for the different degrees of MCD observed in these samples, as broken inversion symmetry is necessary for a finite phonon angular momentum. The $x=0.24$ sample at 12\,K is close to the ferroelectric transition, thus leading to undefined phonon circular motion at low $B$. On the contrary, the samples with higher $x$ are deeper in the ferroelectric phase at the same temperature, and the chiral phonons can be obtained for lower magnetic fields. Additionally, while for the trivial insulator both polarizations maintain similar amplitudes in the studied field range, for the TCI samples the dominant transmitted polarization switches from $E_\text{R}$ to $E_\text{L}$ with increasing $B$. 

In order to quantify the magnetic properties of these chiral phonons, we calculated the real and imaginary parts of the complex Faraday rotation (FR $=\Theta + i \eta$) at a given frequency as $\Theta=[\arg(E_\text{R})-\arg(E_\text{L})]/2$ and $\eta=(|E_\text{L}|-|E_\text{R}|)/(|E_\text{L}|+|E_\text{R}|)$, where $\Theta$ is the (Faraday) rotation of the polarization plane and $\eta$ quantifies the ellipticity change~\cite{xinwei2019}.
%
%

Figures~\ref{fig:3}(a)-\ref{fig:3}(c) present $\Theta$  and $\eta$ for the three samples. For $x=0.24$, the signal again shows multiple wiggles at low $B$, as commented before in relation to data shown in Fig.~\ref{fig:2}(c), but exhibits a pair of positive and negative ellipticity peaks and vanishing Faraday rotation at the same frequency in high magnetic fields. Such evolution for the ellipticity is a direct estimate of the MCD, where chiral phonons will lead to the absorption of either $|E_\text{R}|$ or $|E_\text{L}|$, resulting in a peak of $\eta = \pm1$. As argued before for the MCD in the TCI regime, low magnetic fields are sufficient for the appearance of opposite ellipticity peaks together with vanishing Faraday rotation angles. 


Our findings are consistent with previous FR measurements in Pb$_{0.5}$Sn$_{0.5}$Te~\cite{ChengetAl2019PRL}, which displayed small ellipticity peaks at frequencies similar to ours (although not discussed in their work). In comparison, the lower hole concentrations of our samples allowed us to study the films further into the ferroelectric phase resulting in clearly one chiral sense per phonon. 


The frequencies of the chiral phonons change significantly with the magnetic field strength, as shown in Figs.~\ref{fig:3}(d)-\ref{fig:3}(f). 
The magnetic properties of the observed phonon branches can be extracted using a model for the quadratic dependence in terms of a Zeeman splitting and diamagnetic shift according to $E_\text{L,R}(B) = hf_{\text{TO}} \pm g^* \mu_\mathrm{B} B + \sigma_\mathrm{dia} B^2$, where $h$ is the Planck constant, $f_{\text{TO}}$ is the TO phonon frequency at zero magnetic field, $g^*$ is the effective $g$-factor, $\mu_\mathrm{B}$ is the Bohr magneton, and $\sigma_\mathrm{dia}$ is the diamagnetic shift coefficient~\cite{PhysRevLett.128.075901}. We used this equation in Figs.~\ref{fig:3}(d)-\ref{fig:3}(f), as shown by the solid lines, to obtain the parameters plotted in Figs.~\ref{fig:3}(g)-\ref{fig:3}(i).  

%
%

The trend observed in Fig.~\ref{fig:3}(g) for $f_{\text{TO}}$ agrees with the data presented in Fig.~\ref{fig:1}(e) for the zero-field frequencies at low temperature. Small differences are owing to the large sensitivity of the phonon energies with variations of the nominal temperature in the experiments. Nevertheless, both plots show the same increase of the frequency with $x$, which is a consequence of the composition-dependent phonon hardening, confirming that all films were below the ferroelectric critical temperature during the experiments in magnetic fields. 
For $x=0.24$, the fitting of the Zeeman term for the lowest energy phonon [blue circles in Fig.~\ref{fig:3}(d)] gives a value of $g^*$ ($0.12 \pm 0.02$) that is three times larger than that reported for a $x=0$ sample ($0.043 \pm 0.006$)~\cite{PhysRevLett.128.075901}. The large error bar for $g^*$ in this case is mainly the result of the impossibility to fit the ellipticity peaks down to the low fields. When $x$ is increased into the TCI side, the effective $g$-factor changes sign for TO$_1$ and continues to increase with an opposite sign between the two phonon modes, as shown in Fig.~\ref{fig:3}(h). The value of $g^*$ we obtained for the sample with the largest Sn composition ($g^* = -1.2 \pm 0.2$ for TO$_1$ and $3.3 \pm 0.5$ for TO$_2$) represents a colossal increase, by two orders of magnitude, over the measurements in the trivial insulator samples ($x=0$ and $0.24$). The order of magnitude is comparable with the value recently reported (2.7) for a soft phonon in the Dirac semimetal Cd$_3$As$_2$~\cite{ChengetAl2020NL,PhysRevResearch.1.033101}. 

The diamagnetic term, which increases the energies of both chiral branches with increasing magnetic field, also shows agreement in the order of magnitude between the shift for TO$_1$ in the $x=0.24$ sample and the reported value for $x=0$ [$(1.9 \pm 0.2)\times 10^{-4}$\,meV/T$^2$]. For the same phonon, the diamagnetic shift continuously increases up to one order of magnitude at $x=0.56$ [$(15 \pm 4)\times 10^{-4}$\,meV/T$^2$]. For the TO$_2$ mode, as seen in the Zeeman term, the value is even larger than that for TO$_1$ and reaches $(39 \pm 9)\times 10^{-4}$\,meV/T$^2$.

Because of time reversal symmetry at zero magnetic field, it is expected that each one of the measured phonons (TO$_1$ and TO$_2$) should produce two circularly polarized branches, perhaps with one weaker branch, as reported for PbTe~\cite{PhysRevLett.128.075901}, CeF$_3$~\cite{Schaack_1976}, and Cd$_3$As$_2$~\cite{ChengetAl2020NL}. Nevertheless, due to the MCD, only one type of chirality could be experimentally resolved per phonon mode in Figs.~\ref{fig:3}(d)-\ref{fig:3}(f). Furthermore, those single chiral branches acquired opposite handedness, i.e., left-hand (blue) for TO$_1$ and right-hand (red) for TO$_2$. Using the obtained fitting parameters, we estimated the position of the minority chiral branches, as plotted by the dashed lines. While for $x=0.24$ those energy branches could be in the unidentified FR features, in the TCI regime we expect the R-branch for TO$_1$ to be a shoulder on the low frequency side of the R-branch for TO$_2$ where it is hidden by the peak of the dominant chirality for the other phonon. On the other hand, the L-branch for TO$_2$ should be above its R-branch, which is in the limit or outside the measured frequency window.

Now we discuss possible microscopic mechanisms giving rise to the large effective phonon $g$-factor, $g^*$, and in particular the changes observed across the topological transition. Ren \textit{et al}.~\cite{ren2021phonon} have proposed theoretically a mechanism that leads to phonon effective $g$-factors arising from electronic contributions comparable with the experimental values reported for PbTe~\cite{PhysRevLett.128.075901}. Within this mechanism, valid when the phonon frequencies are smaller than the electronic band gap, the circular motion of a chiral phonon induces an electronic orbital response with topological and nontopological contributions that in turn give rise to a phonon magnetic moment. 

In this work, we adopt a low-energy model for Pb$_{1-x}$Sn$_{x}$Te, taking into account the valley degeneracy and the doping present in the samples, and compute the phonon magnetic moment following Ref.~\cite{ren2021phonon}. We find that this model successfully predicts the increase of the magnetic moment from the case $x=0$ to $x=0.24$ once the occupancy of hole levels in the samples is taken into account~\cite{SM}. The model we employ predicts a magnetic moment sign change across the trivial to TCI phases as observed in our experiments. A $g^*$ sign change was also predicted across a strong-to-weak topological transition~\cite{ren2021phonon}.


However, the model cannot explain the observed continuous enhancement in the TCI phase. Increasing the carrier density decreases $g^*$, contrary to observations~\cite{SM}. A possible explanation is that the surface states contribute to the phonon magnetic moment. However, such a contribution cannot be captured within the premises of our model~\cite{SM}. Recently, Geilhufe proposed a mechanism~\cite{PhysRevResearch.4.L012004} where the chiral phonon induces inertial effects in the electrons leading to an effective spin-chiral phonon coupling that could potentially lead to large phonon magnetic moments. However, the evaluation of this contribution for Pb$_{1-x}$Sn$_{x}$Te requires including the inertial effect Hamiltonian terms in density functional codes. Additionally, the coupling of the phonon to a cyclotron resonance~\cite{ChengetAl2020NL} is negligible as those are largely detuned in the studied range. Further theoretical work is necessary to fully understand the magnetic moment of phonons in the topological phase.

In conclusion, we studied two transverse optical phonon modes in Pb$_{1-x}$Sn$_{x}$Te in a set of thin films across the trivial to topological crystalline insulator transition. We observed the occurrence of a ferroelectric phase in all the samples at a composition-dependent critical temperature. In that phase and under intense magnetic fields, the phonon modes exhibited circular polarization with opposite handedness. While the sample in the trivial side ($x=0.24$) showed magnetic properties in agreement with a previous report for PbTe, the films in the TCI side displayed surprising results. First, a high degree of magnetic circular dichroism could be reached at low magnetic fields. Furthermore, the obtained effective $g$-factors for both phonons increased by two orders of magnitude and changed sign for the lowest-energy phonon mode across the topological transition, thus acquiring opposite signs between the modes. Additionally, the diamagnetic shift showed an increase of one order of magnitude. These results demonstrate that the magnetism of phonons is largely enhanced in topological materials and that a scheme for phonon manipulation with magnetic fields can be more effectively applied in that phase of matter.

We thank Yafei Ren, Di Xiao, Qian Niu, and R.\ Matthias Geilhufe for useful discussions and G.\ Timothy Noe II and Takuma Makihara for assistance in measurements. This research was primarily supported by the National Science Foundation through the Center for Dynamics and Control of Materials: an NSF MRSEC under Cooperative Agreement No.\ DMR-1720595. F.G.G.H.\ acknowledges financial support from the Brasil@Rice Collaborative Grant, the S\~{a}o Paulo Research Foundation (FAPESP) Grants No.\ 2015/16191-5 and No.\ 2018/06142-5, and Grant No.\ 307737/2020-9 of the National Council for Scientific and Technological Development (CNPq). P.H.O.R.\ acknowledges support from CNPq No.\ 307192/2021-0. G.A.F.\ acknowledges additional support from NSF DMR-2114825. J.T.\ and I.K.\ acknowledge support from the Japan Society for the Promotion of Science (JSPS) (KAKENHI No.\ 20H05662).

\bibliography{ref.bib}

\end{document}